\documentclass{PoS}

\newcommand{\ki}{$\chi^2$ }

\newcommand{\etal}{et~al.~}

\newcommand{\kir}{$\chi^2_{\rm red}$}

\newcommand{\sw}{SWIFT~J1753.5$-$0127}
\newcommand{\cyg}{Cygnus~X-1}
\usepackage{wrapfig}

\title{Broad-band spectral changes of the microquasars
Cygnus~X-1 and SWIFT J1753.5$-$0127}

\ShortTitle{Broad-band analysis of Cygnus X-1 and
SWIFT~J1753.5$-$0127}

\author{\speaker{M. Cadolle Bel}\\
        SAp/CEA-Saclay \& AIM,
        Gif-sur-Yvette, France \& ESAC, Madrid, Spain\\
        E-mail: \email{mcadolle@cea.fr}}
\author{M. Rib{\'o}\\
        Universitat de Barcelona, Spain}
\author{J. Rodriguez\\
        SAp/CEA-Saclay \& AIM,
        Gif-sur-Yvette, France}
\author{S. Chaty\\
        SAp/CEA-Saclay \& AIM,
        Gif-sur-Yvette, France}
\author{S. Corbel\\
        SAp/CEA-Saclay \& AIM,
        Gif-sur-Yvette, France} 
\author{A. Goldwurm\\
        SAp/CEA-Saclay \& APC,
        Gif-sur-Yvette, France}

\abstract{We report high-energy results obtained with
\emph{INTEGRAL} and \emph{Rossi-XTE} on two microquasars: the
persistent high-mass system Cygnus X-1 and the transient low-mass
binary SWIFT J1753.5-0127. \emph{INTEGRAL} observed Cygnus X-1
from 2002 to 2004: the spectral (5--1000 keV) properties of the
source, seen at least in three distinct spectral states, show disc
and corona changes. In 2003 June, a high-energy tail at several
hundred keV in excess of the thermal Comptonization model was
observed, suggesting the presence of an additional non-thermal
component. At that time, we detected an unusual correlation
between radio data and high-energy hardness. We also report and
compare the results obtained with simultaneous observations of the
transient source SWIFT J1753.5-0127 performed with
\emph{Rossi-XTE}, \emph{INTEGRAL}, VLA, REM and NTT on 2005 August
10--12 near its hard X-ray outburst. Broad-band spectra and fast
time-variability properties are derived on this source (probably
located in the galactic halo) together with radio, IR and optical
data. We build a spectral energy distribution of the source and
derive interesting multiwavelength constraints. Significantly
detected up to 600 keV in a typical Low/Hard State, the transient
does not seem to follow the usual radio/X-ray correlation.}

\FullConference{VI Microquasar Workshop: Microquasars and Beyond\\
        September 18-22 2006\\
        Societ\`a del Casino, Como, Italy}

\begin{document}

\section{Introduction}

Galactic Black Hole (BH) X-ray binary systems display high-energy
emissions characterized by spectral and flux variabilities (from
milliseconds to months). These systems are found in several
spectral states (see definitions in Homan \& Belloni 2005) giving
us the possibility to study the physical properties of emitting
regions (disc, corona, jets) and their evolutions. Soft and hard
components (and reflection) are coupled to various properties of
variabilities in the power spectrum (e.g., Belloni 2005) and in
the radio (e.g., Corbel et al. 2003). A part from the Low/Hard
States (LHS) and the High/Soft States (HSS) other ones have been
identified characterized either by a greater luminosity than in
the HSS or by variability and X-ray spectral properties mostly
intermediate between the LHS and the HSS: the Hard and Soft
Intermediate States
(HIMS and SIMS).\\
\indent Cygnus~X-1 is one of the first X-ray binaries detected and
has been extensively observed. Among the brightest X-ray sources
of the sky, it is very variable on different time scales and a
relativistic jet has been detected (Stirling et al. 2001). Located
at 2.4$\pm$0.5~kpc, the source accretes by strong stellar wind
from a giant companion. Cygnus~X-1 spends most of its time
(70$\%$) in the LHS. During the SIMS of 1996 June, in addition to
the dominant black body component and the hard component, a
high-energy tail extending up to 10~MeV was discovered (McConnell
et al. 2002).\\
\indent On the other hand, the study of Transient Sources (TS)
which tend to evolve into the LHS in the late stages of an
outburst (e.g., Cadolle Bel \etal 2004) may reveal important clues
on its mechanisms and relation with the accretion rate (e.g.,
Homan et al. 2001; Rodriguez et al. 2003). The TS \sw~was
discovered in hard X rays with the {\it Swift}/Burst Alert
Telescope (BAT) on 2005 May 30 (Palmer et al. 2005). The {\it
Swift}/X-ray Telescope (XRT) observation revealed a variable
source; the source was also clearly detected in UV with the
UV/optical Telescope (Still \etal 2005). On the ground, the
optical MDM 2.4~m telescope revealed a new star within the {\it
Swift} error circle (Halpern et al. 2005). At the beginning of
July, Fender \etal (2005) reported with MERLIN a probable
point-like radio counterpart (consistent with a compact jet). In X
rays, the 1.2--12~keV source flux increased to the maximum value
of $\sim$~200 mCrab in few days and started to decay slowly. The
hard power law spectrum observed with the {\it Swift}/XRT (Morris
\etal 2005) and the 0.6~Hz QPO detected in pointed {\it RXTE}
observations (Morgan \etal 2005) are characteristic of the LHS as
also reported later by Miller \etal (2006).\\
\indent We report exciting results collected on \cyg~over two
years with {\it INTEGRAL} (e.g., Cadolle Bel et al. 2006a,
hereafter CB06a; Malzac et al. 2006). In parallel, we also report
the results of our triggered Target of Opportunity (ToO) campaign
for TS in the galactic halo (August 10--12) on \sw. Preliminary
results reported in Cadolle Bel \etal (2005) showed that the TS
was then still in a LHS. Thanks to a large multi-wavelength
program we could also trigger simultaneously optical, infrared and
radio ToOs and we added \emph{RXTE} data in our analysis as
explained in
Cadolle Bel et al. 2006b (hereafter CB06b).\\

\section{Observations and data reduction}

\subsection{Cygnus X-1}
\begin{figure}[htbp]
\begin{minipage}[b]{0.49\linewidth}
\centering \includegraphics[width=7cm]{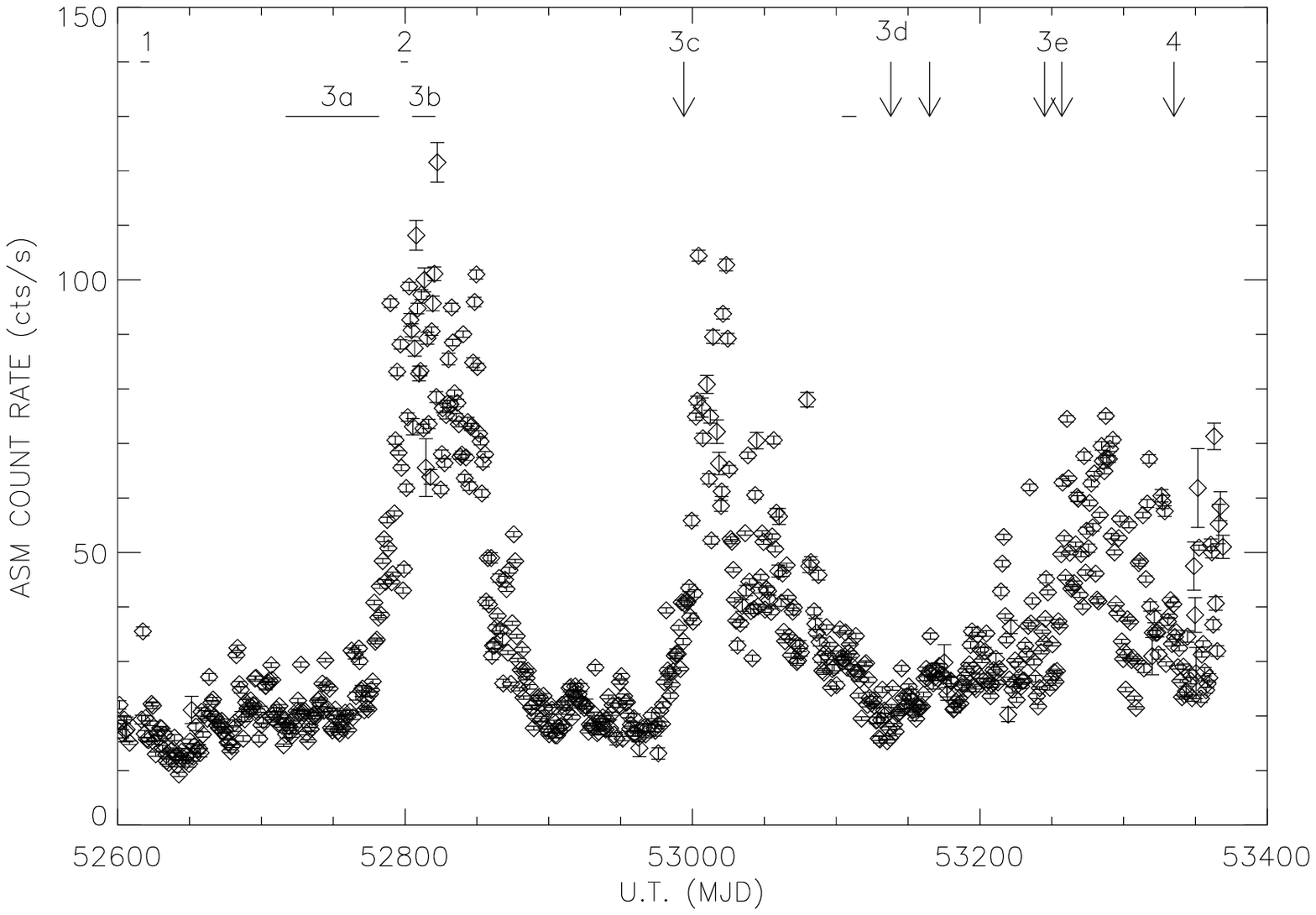}
\end{minipage} \hspace{0.2cm}
\begin{minipage}[b]{0.49\linewidth}
\centering \includegraphics[width=7cm]{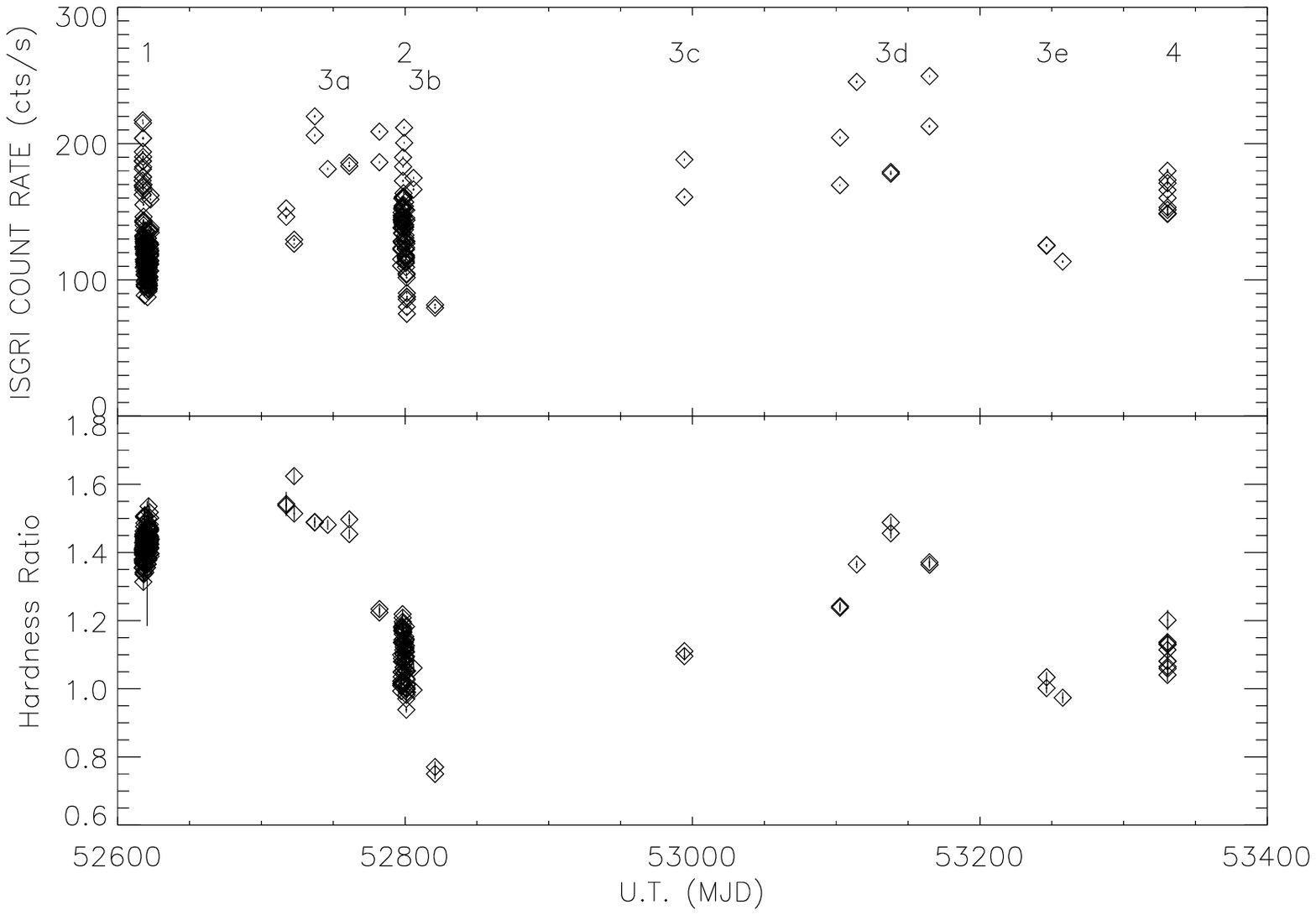}
\end{minipage}
\caption{\label{LCHR} \emph{Left}: \emph{RXTE}/ASM daily average
(1.2--12~keV) LC of \cyg~from 2002 November to 2004 November with
the period of our \emph{INTEGRAL} observations (see text and
Table~1 for epoch definitions). \emph{Right}: The 20--200 keV
IBIS/ISGRI LC of \cyg~(top) and hardness ratio (bottom) between
the 40--100 keV and 20--30 keV energy bands.}
\end{figure}
\indent The periods of our {\it INTEGRAL} observations (epochs 1
to 4) are indicated on Fig.~\ref{LCHR} (left) on the
\emph{RXTE}/ASM Light Curve (LC). To discuss the time evolution of
the source, IBIS/ISGRI LC and Hardness Ratios (HR) obtained over
two years are reported in Fig.~\ref{LCHR} (right). Epoch 1 (2002
December 9--11) includes part of the PV-Phase observations of
Cygnus X-1. Epoch 2 corresponds to an Open Time observation
performed on 2003 June 7--11 while epochs 3 and 4 refer
respectively to the set of Cygnus~X-1 observations during the GPS
and the 2004 November calibrations.\\

\subsection{SWIFT J1753.5-0127}
\begin{figure}[htbp]
\begin{minipage}[b]{0.49\linewidth}
\centering \includegraphics[width=7cm]{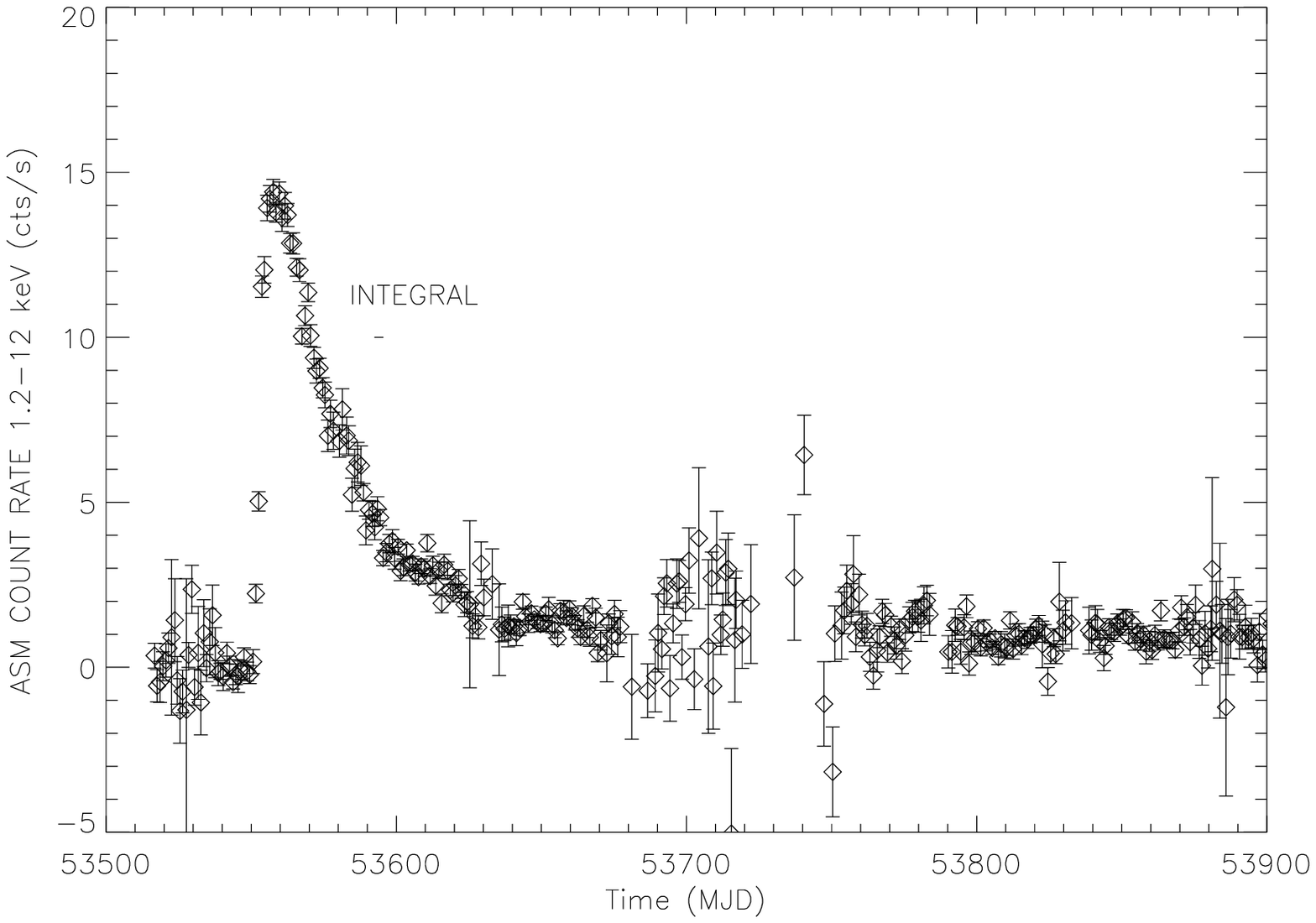}
\end{minipage}\hspace{0.2cm}
\begin{minipage}[b]{0.49\linewidth}
\centering \includegraphics[width=7cm]{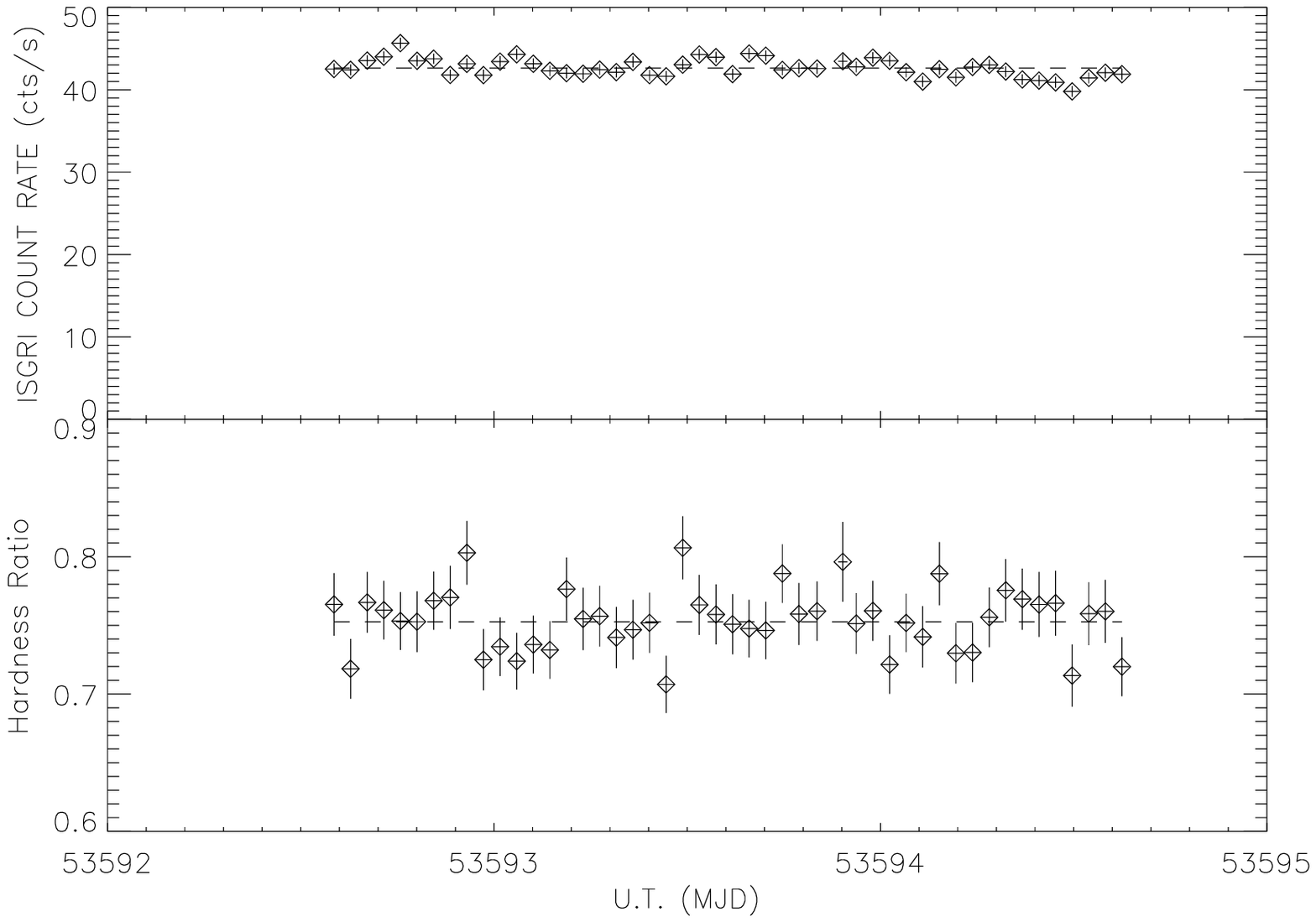}
\end{minipage}
\caption{\label{LCHRsw} \emph{Left}: \emph{RXTE}/ASM daily average
(1.2--12~keV) LC of \sw~from 2005 mid-May up to 2006 mid-June with
our \emph{INTEGRAL} ToO indicated. \emph{Right}: The 20--320 keV
IBIS/ISGRI LC of \sw~(top) and corresponding HR (bottom) between
the 40--80 and 20--40 keV energy bands (errors at the 90$\%$
confidence level).}
\end{figure}
\indent The journal of our simultaneous multiwavelength
observations to our \emph{INTEGRAL} ToO is given in CB06b in which
we also detail our analysis procedures. Fig.~\ref{LCHRsw} (left)
plots the 1.2--12~keV \emph{RXTE}/ASM daily average LC of
\sw~since the discovery of the TS up to 2006 June 15 while
Fig.~\ref{LCHRsw} (right) shows the IBIS/ISGRI LC and HR during
our multiwavelength ToO. Besides on 2005 August 11 (around UT 02)
we also obtained optical photometry in $B$, $V$ $R$ and $I$ bands
with the spectro-imager EMMI (NTT) and we observed \sw~with the
NRAO VLA at 1.4, 4.9, 8.5 and 15~GHz on 2005 August 11 (average
MJD~53593.28) with
the VLA.\\

\section{Spectral results on Cygnus X-1}

\indent As shown in Fig.~\ref{LCHR} (left), during the epoch 2
{\it INTEGRAL} observations, the 1.5--12~keV ASM average count
rate of Cygnus~X-1 ($\sim$~1.3 Crab) was larger than during epoch
1 ($\sim$~290 mCrab) by a factor of 4.5. The derived IBIS/ISGRI
20--200 keV LC and HR of Cygnus~X-1 are shown in Fig.~\ref{LCHR}
(right, epochs 1 to 4). From epoch 1 to epoch 2, while the ASM
average count rate increased, the 20--200 keV IBIS/ISGRI one
decreased from $\sim$~910 to $\sim$~670~mCrab. This probably
indicates a state transition as also suggested by the decrease in
the IBIS HR (the source softens). Similar transitions, with a
change in the ASM LC and an evolving IBIS HR, occurred again
during epoch 3 and epoch 4. Table~\ref{tab:para} gives all the
best-fit parameters of \cyg~with a model involving Comptonization
(Titarchuk 1994), reflection (Magdziarz \& Zdziarksi 1995) and
when needed a multicolor black body disc (Mitsuda et al. 1984) and
Fe line components (modelling
approach described in CB06a).\\

\begin{table*}[htbp]
\begin{center}
\caption{\label{tab:para}~Best-fit parameters of Cygnus~X-1 for
the current thermal model in the several observation epochs. Model
in {\tt XSPEC} notations: {\sc constant}*{\sc wabs}*({\sc
diskbb}+{\sc gaussian}+{\sc reflect}*{\sc comptt}) with $N_{\rm
H}$=6~$\times$~10$^{21}$cm$^{-2}$ and $kT_{\rm 0}$ value tied to
disc $kT_{\rm in}$. Errors are at 90$\%$ confidence level
($\Delta$\ki~=~2.7).}
\begin{tabular}[h]{llllllll}
\hline Epochs, &~~~~$K$$^a$ &$kT_{\rm in}$ or $kT_{\rm
0}$&~~$kT_{\rm e}$ &~~~~~~$\tau$ &~~~$E_{\rm Fe}$
line&~~~$\Omega/2\pi^b$&~~~~\kir \\
dates (MJD)&&~~~~(keV)&~(keV) & &~~~~(keV)&&~~~(dof)\\
\hline
1, 52617-620 &~~~~-&0.20 (frozen) &67$_{-6}^{+8}$&1.98$_{-0.23}^{+0.21}$&~~~~~~~-&0.25$_{-0.04}^{+0.03}$&1.45 (230)\\
2, 52797-801&250$_{-59}^{+89}$&1.16$\pm$0.07&100$_{-17}^{+29}$&0.98$_{-0.28}^{+0.25}$&7.07$_{-0.11}^{+0.12}$&0.57$_{-0.06}^{+0.09}$&1.69 (236) \\
3{\it{a}}, 52710-780&~~~~-&0.20 (frozen)&68$_{-12}^{+22}$&2.08$_{-0.84}^{+0.51}$&6.48$\pm$0.13&0.32$_{-0.07}^{+0.05}$&1.07 (190) \\
3{\it{b}}, 52801-825&312$_{-24}^{+25}$&1.15$\pm$0.03&93$\pm$42&0.80$~_{-0.40}^{+0.86}$&6.40$\pm$0.73&0.58$_{-0.18}^{+0.20}$&0.93 (190)\\
3{\it{c}}, 52990&361$_{-67}^{+61}$&0.99$\pm$0.08&58$_{-15}^{+54}$&1.60$_{-0.80}^{+0.64}$&6.96$\pm$0.19&0.23$_{-0.09}^{+0.17}$&0.99 (190)\\
3{\it{d}}, 53101-165&~~~~-&0.20 (frozen)&56$_{-7}^{+12}$&2.28$_{-0.41}^{+0.30}$&6.11$\pm$0.26&0.27$\pm$0.06&0.81 (190)\\
3{\it{e}}, 53240-260&132$\pm$10&1.39$\pm$0.77&48$_{-6}^{+20}$&1.85$_{-0.07}^{+0.40}$&6.49$\pm$0.38&0.49$_{-0.32}^{+0.37}$&1.56 (190)\\
4, 53335&232$_{-32}^{+21}$&1.16 (frozen)&128$_{-63}^{+84}$&0.74$_{-0.38}^{+0.88}$&7.78$_{-0.42}^{+0.44}$&0.47$_{-0.14}^{+0.18}$&0.97 (221)\\
\hline
\end{tabular}
\end{center}
Notes:\\
~a)~Disc normalization $K~=~(R/D)^{2}~\cos~\theta$ ($R$: inner
disc radius in units of km; $D$: distance to the source
in units of 10 kpc; $\theta=45^\circ$).\\
~b)~Solid angle of the reflection component.\\
\end{table*}

\begin{figure}[htbp]
\begin{minipage}[b]{0.49\linewidth}
\centering \includegraphics[width=7cm]{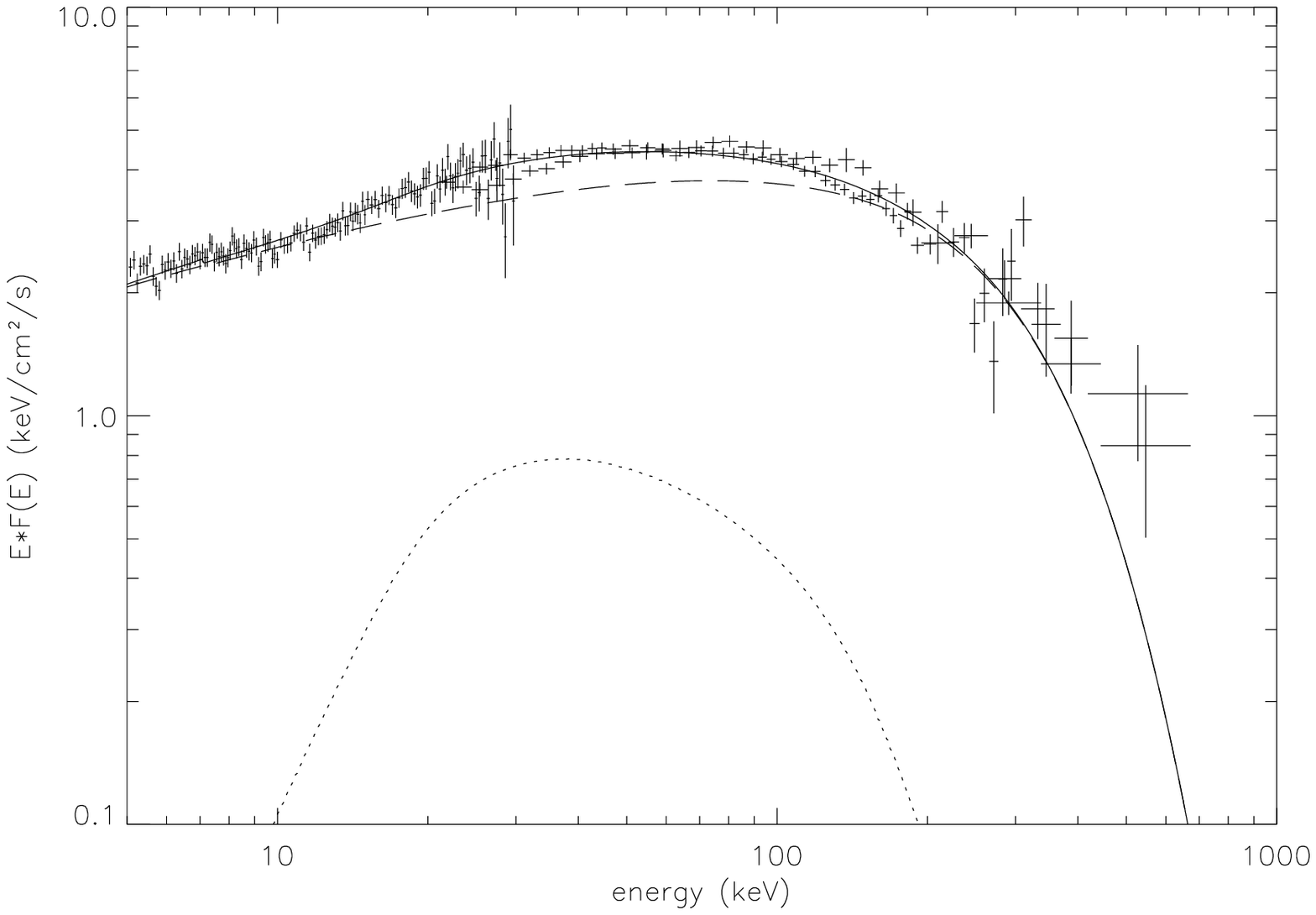}
\end{minipage}\hspace{0.2cm}
\begin{minipage}[b]{0.49\linewidth}
\centering \includegraphics[width=7cm]{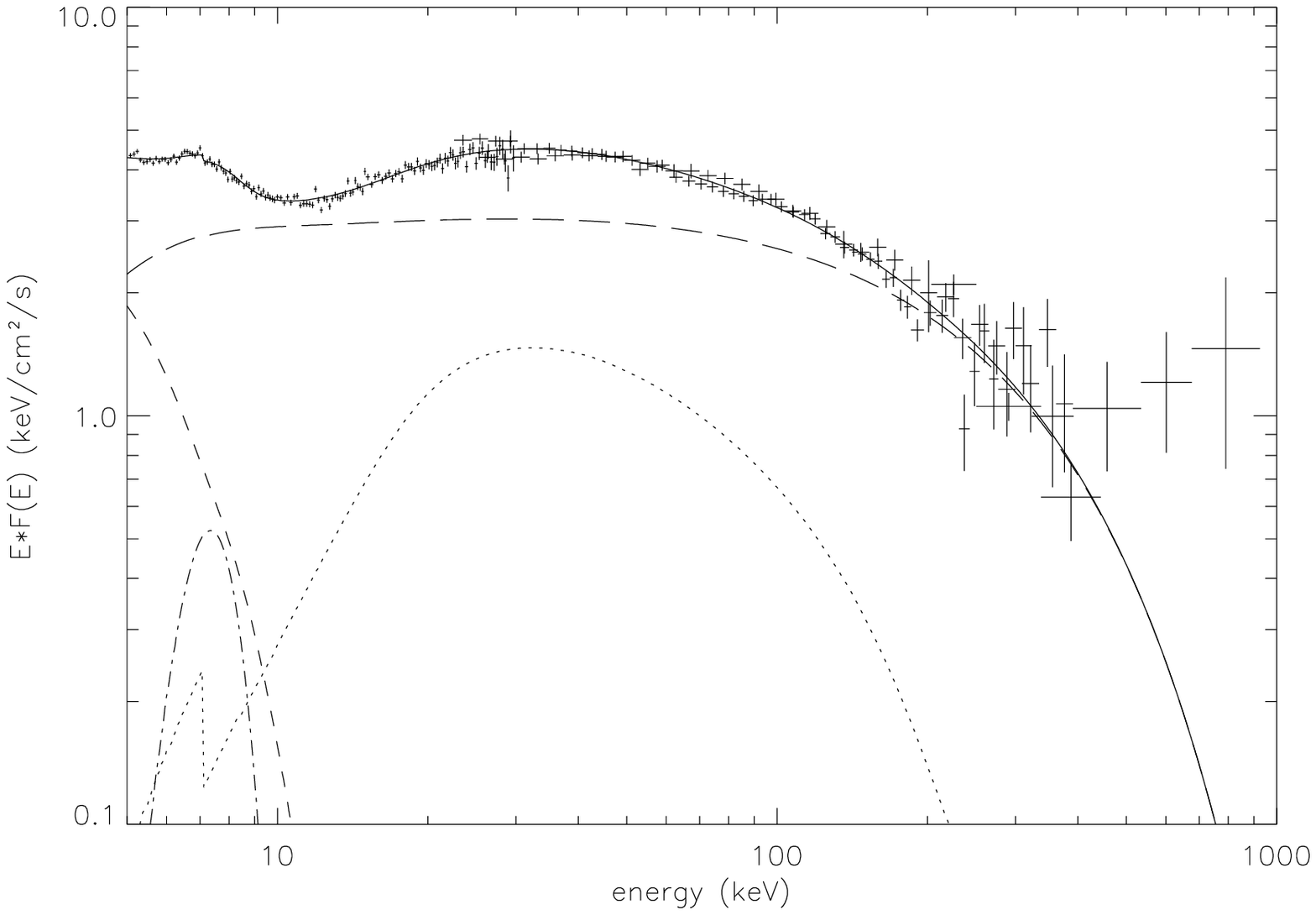}
\end{minipage}
\caption{\label{spec}{{\it{Left:}} Epoch 1 unabsorbed $EF(E)$
spectrum of \cyg~along with the best-fit model described in
Table~1 with the JEM-X, SPI and IBIS (ISGRI and PICsIT) data.
{\it{Right:}} The same for epoch 2. \emph{Dotted}: reflection.
\emph{Long dashes}: Comptonization. \emph{Dashed}: disc.
\emph{Dotted-dashed}: gaussian line. \emph{Thick}: total model.}}
\end{figure}

\subsection{The Low/Hard State spectrum}

\indent Fig.~\ref{spec} (left) shows the resultant $EF(E)$
spectrum and its best-fit with the JEM-X, IBIS and SPI data during
epoch 1. The best-fit model is reported in Table~\ref{tab:para}.
The disc black body is very weak or below the energy range of
JEM-X. While the 20--100~keV luminosity is
6.5~$\times$~10$^{36}$~erg~s$^{-1}$ (at 2.4~kpc), the bolometric
(extrapolated from 0.01~keV to 10~MeV) luminosity has the value of
2.2~$\times$~10$^{37}$~erg~s$^{-1}$. These parameters are
consistent with those found in BH binaries in the LHS as fully
discussed in CB06a.\\

\subsection{Transitions to Soft Intermediate States (SIMS)}

\begin{figure}[htbp]
\begin{minipage}[b]{0.49\linewidth}
\centering
\includegraphics[width=6cm]{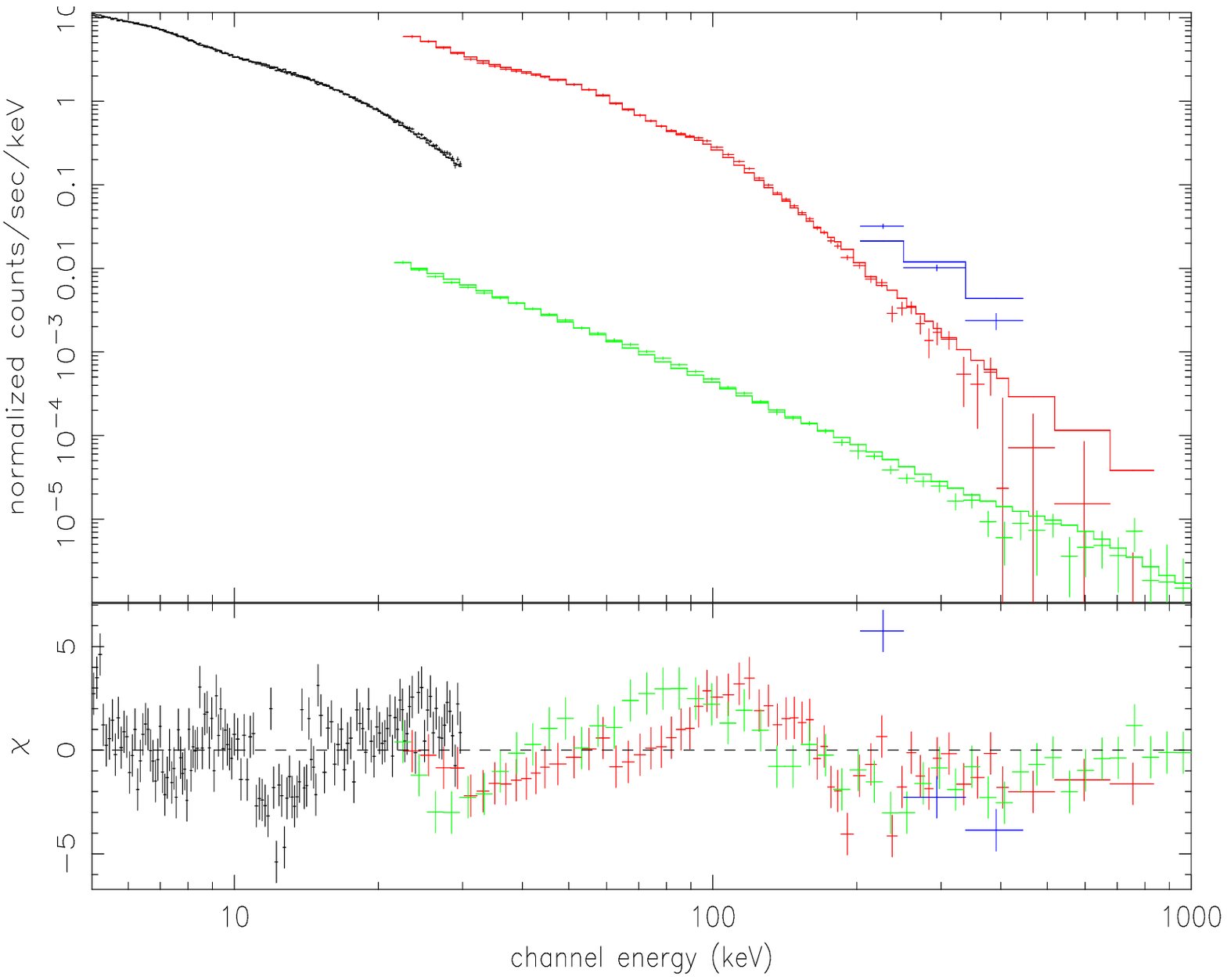}
\end{minipage}\hspace{0.2cm}
\begin{minipage}[b]{0.49\linewidth}
\centering \includegraphics[width=6cm]{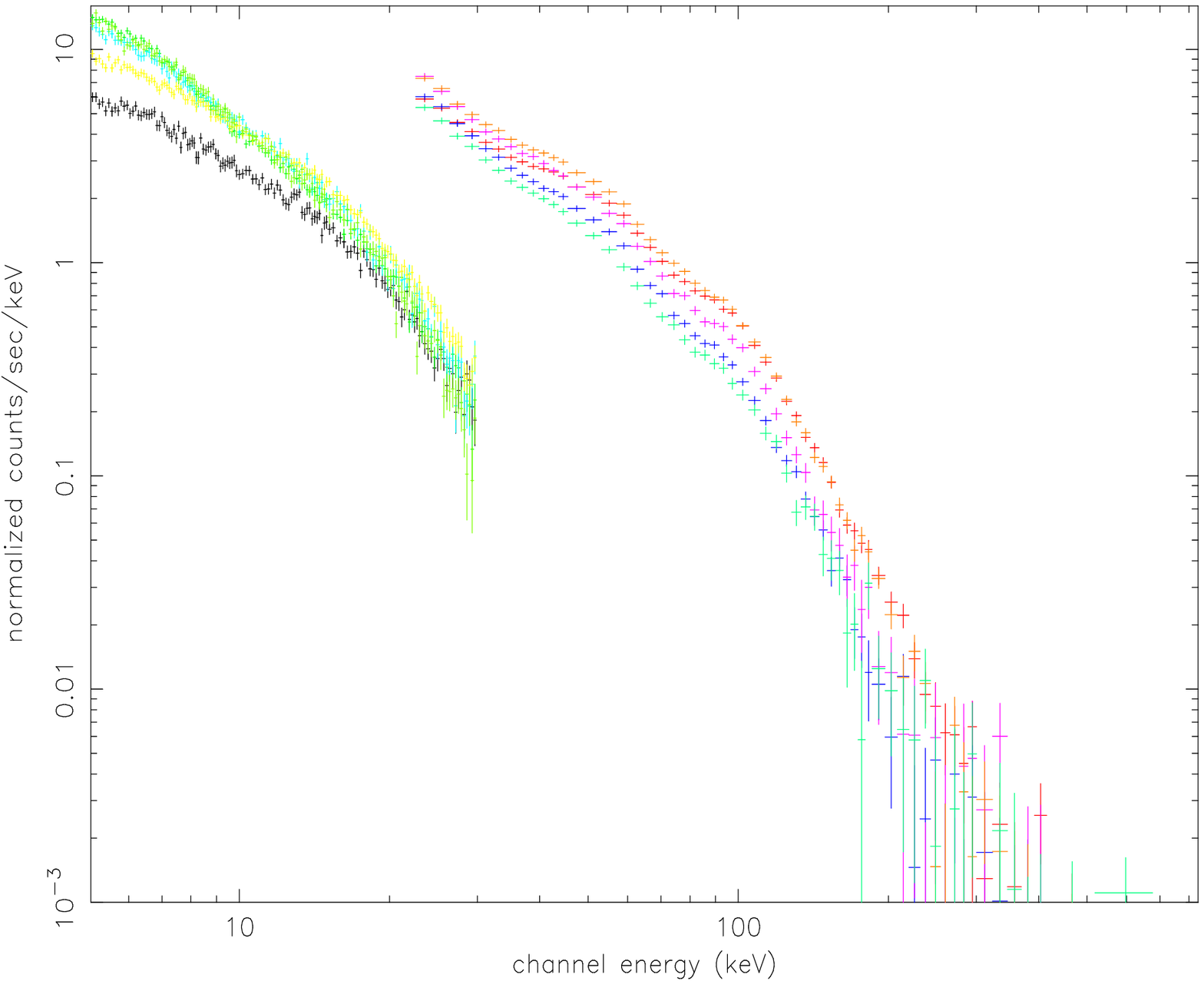}
\end{minipage}
\caption{\label{eqpair} \emph{Left}: Spectra of \cyg~during epoch
2 with JEM-X (black), SPI (green) and IBIS (ISGRI: red; PICsIT:
blue) along with the best-fit hybrid model of Coppi (1999)
including a non-thermal distribution for corona particles.
Residuals in $\sigma$ units are also shown. \emph{Right}: Spectra
of \cyg~during epoch 3 with JEM-X and IBIS/ISGRI along with the
best-fit hybrid model (see Table~1).}
\end{figure}
Fig.~\ref{spec} (right) shows the resultant $EF(E)$ spectrum and
its best-fit with the JEM-X, IBIS and SPI data during epoch 2.
Table~\ref{tab:para} summarizes the best-fit parameters and the
\kir~obtained from 5~keV up to 1~MeV. We get a plasma temperature
and an optical depth respectively higher and lower than in epoch
1.
The disc accounts for 26~$\%$ of the total luminosity and the
reflection is higher in epoch 2 than in epoch 1. Considering the
behaviour of the ASM, IBIS LC and HR (Fig.\ref{LCHR}), the
relative softness of the spectrum and the presence of a relatively
strong hard energy emission, it appears that during the 2003 June
observations Cygnus~X-1 was in the SIMS. This is also confirmed by
radio observations of Malzac \etal (2006) who suggested that the
fluctuations of the radio luminosity were associated with a
pivoting of the high-energy spectrum and that the source did not
display the usual radio/X-ray correlation. The derived thermal
Comptonization parameters are consistent with those found in BH
binaries in SIMS (McClintock \& Remillard 2006).\\
\indent As one can be seen in Fig.~\ref{spec} (right), an excess
with respect to the Comptonized spectrum above 400~keV is observed
in the SPI data (not present in epoch 1 and not due to
instrumental effects). Consequently we fitted the data with the
hybrid model of Coppi (1999) coupled to the usual disc and Fe line
components. Fig.\ref{eqpair} (left) shows the resultant count
spectrum obtained in epoch 2 with this model: with a \kir~=~1.55
(232 dof), clearly better than the current epoch 2 thermal model,
the derived thermal values of $\tau$, $\Omega/2\pi$, $E_{\rm Fe}$
centroid and EW match, within the uncertainties, the parameters
obtained in Table~\ref{tab:para}. The value of $kT_{\rm e}$
(42~keV) decreases from the pure thermal model as expected. The
non-thermal electron power represents $\sim$~16$\%$ of the total
power supplied to the electrons in the corona. The inferred
bolometric luminosity is 3.3~$\times$~10$^{37}$~erg~s$^{-1}$.
Similar spectral transitions seem to occur later: epochs 3
{\it{a}} to {\it{e}}) are close pointings which occur (see
Fig.~\ref{LCHR}) in different regimes of ASM count rate and of
average IBIS HR. The best-fit spectral results
(Table~\ref{tab:para}) we obtained on JEM-X and IBIS/ISGRI data
(Fig.~\ref{eqpair}, right) indicate that, during sub-groups 3{\it
a} and {\it d}, Cygnus~X-1 was in a LHS (as in epoch 1) while, in
sub-groups 3{\it b}, {\it c} and {\it e} and in epoch 4, the
source was in a softer state (HSS or SIMS) as explained in CB06a.\\

\section{Multiwavelength results on SWIFT J1753.5-0127}

\subsection{Light curves and timing variabilities}

\indent From the end of 2005 May up to July 9, the ASM average
count rate increased (Fig.~\ref{LCHRsw}, left): its flux reached
the maximum value of $\sim$ 200~mCrab (MJD~53560) and then
decreased to $\sim$ 14~mCrab (MJD~53650). The characteristic decay
time we derived (37.0~$\pm$~0.2 days) is compatible with the usual
behaviour of TS in outburst (Tanaka \& Shibazaki, 1996; Chen et
al. 1997) like, e.g., XTE~J1720$-$318 (Cadolle Bel \etal 2004).
During our {\it INTEGRAL} ToO the IBIS/ISGRI count rate was almost
constant at 43~cts~s$^{-1}$ ($\sim$ 205~mCrab) between MJD
53592--53594.4 (Fig.~\ref{LCHRsw}, right, top) with a constant HR
($\sim$0.75).\\
\indent We produced a Power Density Spectrum (PDS) with the PCA
data : the continuum is well represented by the sum of two
zero-centered broad Lorentzians while an additional third
Lorentzian is needed to account for a QPO around 0.24~Hz. This
value is lower than the 0.6~Hz QPO reported after the peak of the
outburst: this trend is sometimes observed in other BHCs and has
been associated to the recession of the accretion disc (Kalemci
\etal 2002; Rodriguez \etal 2004; Belloni \etal 2005). However, we
can not exclude another interpretation for the QPO based on the
pulsation modes of the corona (Shaposhnikov \& Titarchuk 2006). In
any case the high level of band-limited noise ($\sim$ 27$\%$
r.m.s.) we observed is
typical of the LHS.\\

\subsection{X/$\gamma$-ray spectral results and constraints}

\begin{figure}[htbp]
\begin{minipage}[b]{0.49\linewidth}
\centering
\includegraphics[width=7cm]{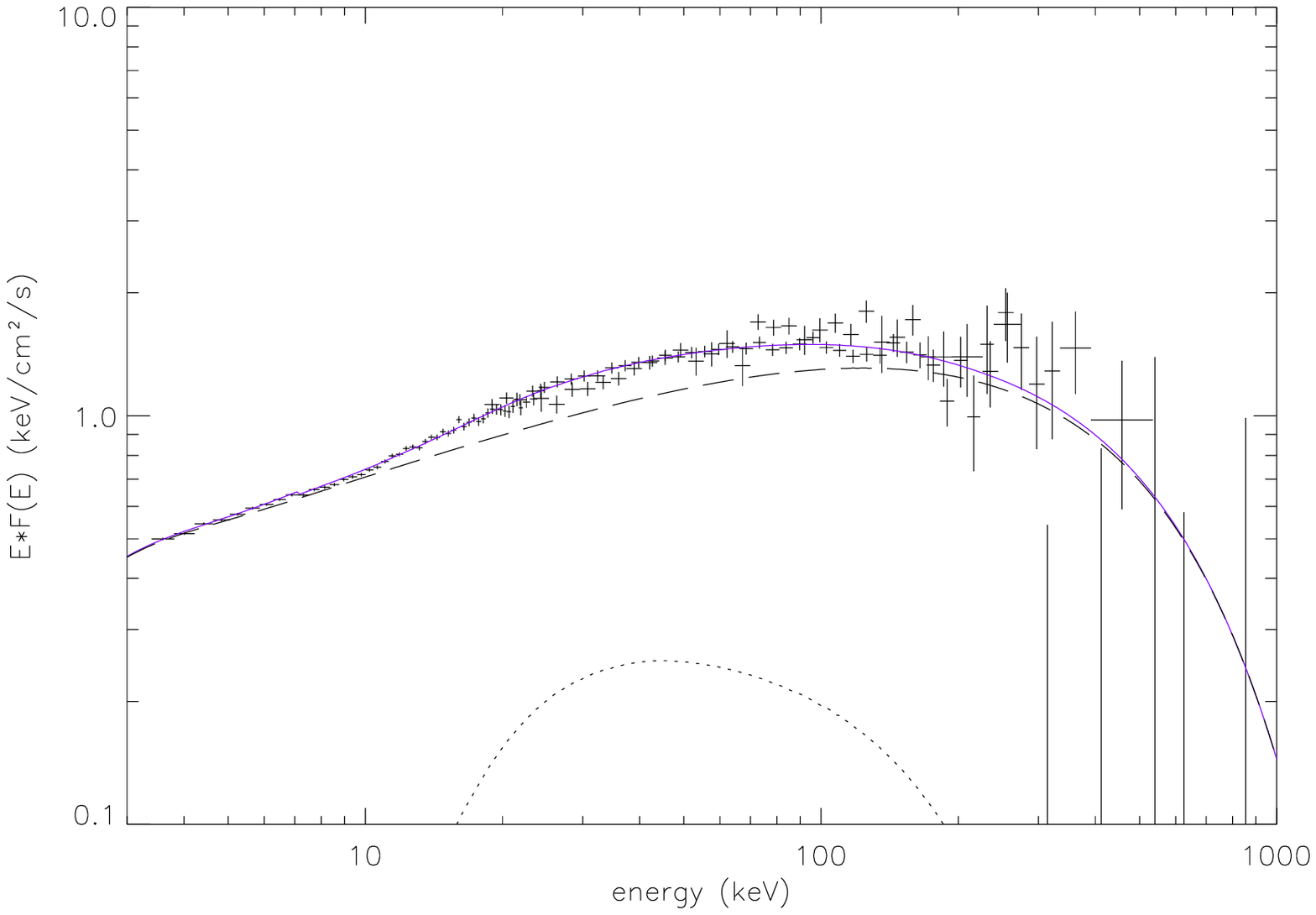}
\end{minipage}\hspace{0.2cm}
\begin{minipage}[b]{0.49\linewidth}
\centering \includegraphics[width=7cm]{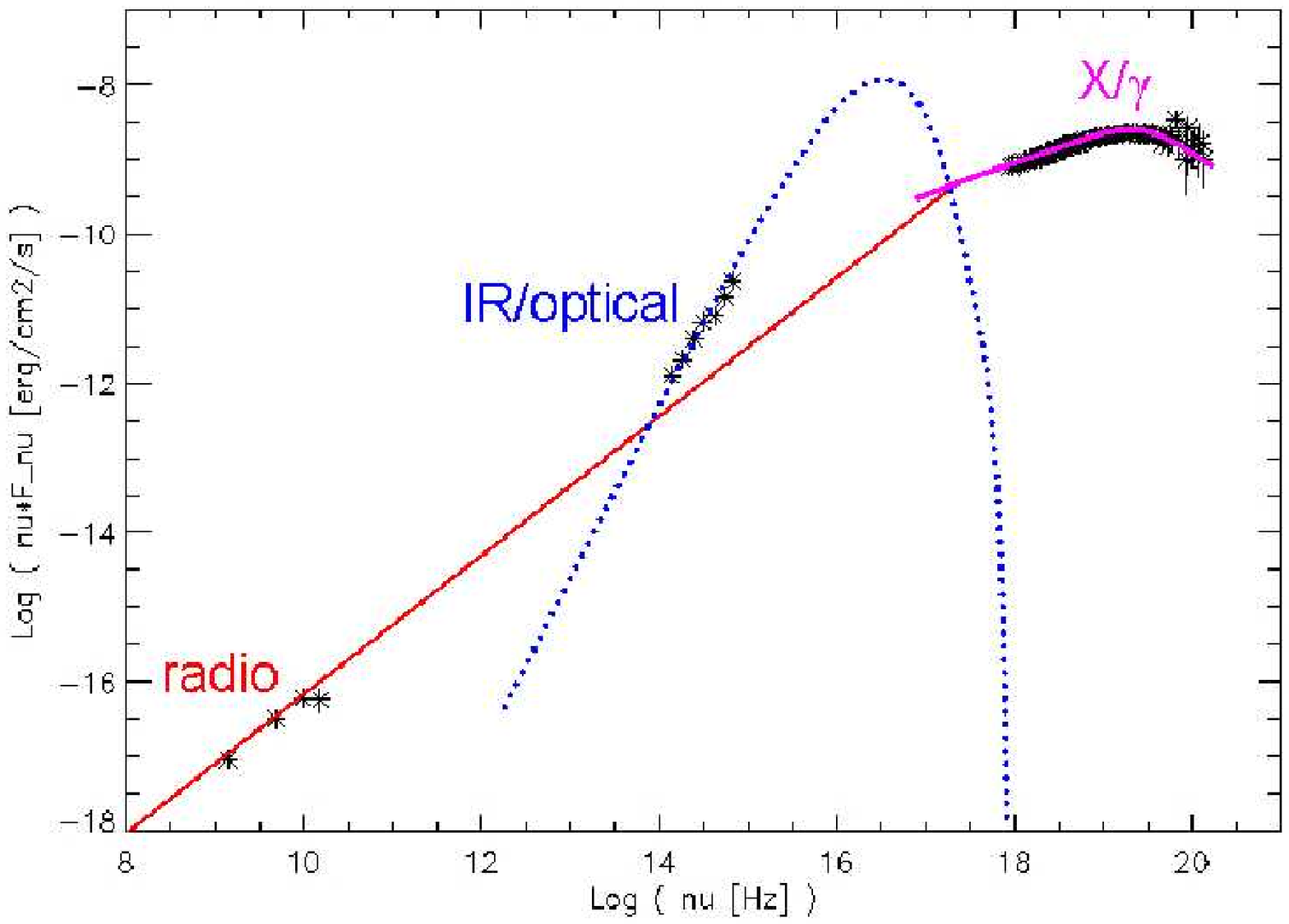}
\end{minipage}
\caption{\label{tot} \emph{Left}: $EF(E)$ spectra of \sw~during
our {\it INTEGRAL} ToO with the PCA, IBIS and SPI data along with
the best-fit model (thick): absorbed Comptonization (dashed)
convolved by reflection (dotted). \emph{Right}: SED of \sw~on
August 11 with radio, IR, optical and X-ray points (up to 535~keV)
and errors. The IR and optical flux densities were dereddened (see
CB06b). At least three distinct contributions are seen and
interpreted as the optically thick synchrotron emission from a
compact jet (red); a thermal disc (blue) and the Comptonization of
soft photons by a hot medium (pink).}
\end{figure}
\indent As there is no significant variation in the HR
(Fig.~\ref{LCHRsw}, right, bottom), we therefore used the whole
data from JEM-X, IBIS/ISGRI and SPI of this hard outburst to build
up an average spectrum on a wide band together with the
simultaneous PCA and HEXTE data. Following the approach described
in CB06a when modelling the LHS spectra of Cygnus~X-1 and in
CB06b, with the {\sc cons}*{\sc wabs}*({\sc reflect}*{\sc comptt})
model (in XSPEC notation), we obtain a reasonable \kir~of 1.17
(with 121 dof) and $kT_{\rm 0}$=0.54$_{-0.07}^{+0.04}$~keV,
$kT_{\rm e}$=150$\pm$26~keV, $\tau$=1.06$\pm$0.02 with
$\Omega/2\pi$=0.32$\pm$0.03. The relatively high Comptonization
temperature can be interpreted as the presence of a medium
(corona) which remains hot because of a less important cooling
from a reduced number of soft disc photons. The best-fit model
over-plotted on the data is reported on Fig.~\ref{tot} (left) in
$EF(E)$ units. The parameters we derive are compatible with the
source being in the LHS (e.g., Cadolle Bel \etal 2005, CB06b). The
Comptonization parameter $y$ ($=kT_{\rm e}/m_{\rm e}c^2$~Max
($\tau$, $\tau^2$)) is typical of a LHS ($\sim$0,33) as fully
discussed in CB06a for Cygnus~X-1.\\
\indent While our data start at 3~keV, leading to a possible
underestimation of the bolometric luminosity, we derive an
unabsorbed 2--11~keV flux of $1.5 \times
10^{-9}$~erg~cm$^{-2}$~s$^{-1}$ and a bolometric flux of 1.3$
\times 10^{-8}$~erg~cm$^{-2}$~s$^{-1}$. This corresponds to an
unabsorbed bolometric luminosity of 5.77~$(d/6~{\rm
kpc})^2$$\times$~10$^{37}$~erg~s$^{-1}$ well below the Eddington
regime (even for a low BH mass of 1~M$_\odot$). Computing the
bolometric luminosity for different distances to \sw~we derive
minimum compact object masses to guarantee that this corresponds
to less than 5$\%$ of the Eddington luminosity, as seen for BH in
the LHS (Maccarone et al. 2003): a 3 M$_\odot$ BH implies at least
a distance of 4 kpc.

\subsection{Optical and radio results}

Comparisons of the spectra obtained between July (Torres et al.
2005a, b) and our ToO show the expected behaviour of LMXBs in
outburst: bright contribution of a disc in optical followed by a
decrease of this contribution simultaneously to a decrease of the
soft X-ray flux. We could determine a column density of
(1.97$\pm$0.23)$\times$10$^{21}$~atoms~cm$^{-2}$ along the line of
sight: this value is consistent with the absorption determined by
{\it Swift}/XRT and it would place the source at $\sim$6~kpc (to
be compatible with its absorption and its high latitude value of
$l$=12.9$^\circ$) without requiring intrinsic absorption, with a
similar height above the Galactic Plane than those of the LMXB
halo sources XTE~J1118+480 and Scorpius~X-1. Considering the
angular resolution of the VLA (the synthesized beam), we detected
at all observed frequencies a point-like radio counterpart
(angular radius $<4^{\prime\prime}$) at a position compatible with
the MERLIN one. The radio data are nearly compatible with a flat
-or slightly inverted- spectrum since the best-fit leads to
$\alpha$=$-$0.17$\pm$0.16 (where $S_\nu\propto\nu^{+\alpha}$).

\subsection{Spectral Energy Distribution (SED)}

\indent We calculated for radio, IR (with REM-ROSS J, H and K
bands thanks to P. D'Avanzo), optical and X-ray data the
corresponding flux value in $\nu F_\nu$ units corrected for
extinction. The full procedure is described in CB06b. We show in
Fig.~\ref{tot} (right) the SED of \sw~from radio to X rays in a
logarithmic scale. The SED reveals that at least three distinct
contributions are necessary: its shape is similar to the ones
observed for transient LMXBs (e.g., XTE~J1118$+$480, Chaty \etal
2003;
XTE~J1720$-$318, Chaty \& Bessolaz 2006).\\

\subsection{Optical and radio constraints}

\indent Comparing with nearby faint USNO-B1.0 stars (Monet \etal
2003), we estimate a quiescent visual magnitude above 19.5~mag
(from $R > 19.0\pm0.5$ and $B > 20.0\pm0.5$). Using for example
the absolute visual magnitudes from Ruelas-Mayorga (1991), even
for the less luminous intermediate type giant companions in the
range F8-G2\,III, the distance to the source should be
$\sim$15~kpc, implying a very high minimum BH mass of
$\sim$55~$M_\odot$ to guarantee $L_{\rm bol}<5\%\ L_{\rm Edd}$.
Clearly, an intrinsically fainter donor is required (a main
sequence type K or M companion) rather than earliest types,
ranging \sw~in the LMXB class. Besides, the flat radio spectrum of
\sw~is similar to the ones typically found in BH during LHS (e.g.,
Fender \etal 2005): it is usually interpreted as synchrotron
radiation produced in a partially self-absorbed conical and
compact jet (Gallo \& Gallo et al. 2003). It is not resolved in
our data because it is too faint.\\

\subsection{Discrepancy with the radio/X-ray correlation}

\indent Corbel \etal (2003) then Gallo \etal (2003) found a
correlation between the X-ray flux and the radio flux density for
BH in the LHS (scaled to 1~kpc). We have used our measured
unabsorbed X-ray flux to compute the expected radio flux density
according to their correlation by using different possible
distances to \sw: the measured value is one order of magnitude
lower than the expected one, even for the highest possible
distances to the source. This behaviour was already observed for,
e.g., XTE~J1650$-$500 (Corbel \etal 2003) and implies that we
probably do not constrain very well the $k$ value.\\

\section{Discussion}

\indent In these two microquasars, we have observed distinct
spectral states and we have determined interesting multiwavelength
constraints. Using the broad-band capability of {\it INTEGRAL}, it
has been possible to accumulate a large amount of data on
Cygnus~X-1 between 5~keV--1~MeV to follow its spectral evolution
from 2002 to 2004. We characterized Comptonization parameters
changes of the source correlated to the presence of a variable
disc emission indicating transitions between the LHS and softer
(Intermediate) states. Besides, a high-energy tail during the SIMS
emerged from pure Comptonization between 400~keV--1~MeV and was
probably associated with a non-thermal component. Also,
unusual radio/X-ray correlation was detected.\\
\indent Besides, we have accurately studied \sw~over a wide energy
band (3~keV--1~MeV). While Comptonization fits well our
X/$\gamma$-ray data from \emph{RXTE} and \emph{INTEGRAL} we found
that, although clearly in LHS, this source is interestingly well
below the radio/X-ray correlation (even assuming a large
distance). Another possibility is that \sw~should radiate more
than the Eddington regime but this has never been observed before
for a BH in the LHS. The emission could be compatible with the
standard picture of synchrotron and inverse Compton radiation
coming from a self-absorbed conical jet (Markoff et al. 2005). The
extent to which the spectrum hardens at energies approaching 1~MeV
has now become an important issue for theoretical modelling of the
accretion processes and radiation mechanisms in BH binaries and
TS. Such studies will shed light on the accretion processes and
radiation mechanisms at work in their
vicinity. \\


\begin{thebibliography}{99}
\bibitem{} Belloni, T. 2005, astro-ph 0507566
\bibitem{} Cadolle Bel, M., Rodriguez, J., Sizun, P., \etal 2004, A\&A, 426, 659
\bibitem{} Cadolle Bel, M., Rodriguez, J., Goldwurm, A., \etal 2005, ATEL 574
\bibitem{} Cadolle Bel, M., Sizun, P., Goldwurm, A., et al. 2006a,
A\&A, 446, 591
\bibitem{} Cadolle Bel, M., Rib\'o, M., Rodriguez, J., et al. 2006b,
submitted to ApJ
\bibitem{} Coppi, P.S., 1999, ASPC, 161, 375
\bibitem{} Chaty, S., Haswell, C. A., Malzac, J.,
\etal 2003, MNRAS, 346, 689
\bibitem{} Chaty, S. \& Bessolaz, N. 2006,
accepted for publication in A\&A, astro-ph 0605297
\bibitem{} Chen, W., Shrader, C. R. \& Livio, M. 1997, ApJ, 491, 312
\bibitem{} Corbel, S., Nowak, M. A., Fender, R. P., et al. 2003,
A\&A, 400, 1007
\bibitem{} Corbel, S., Fender, R. P., Tomsick, J. A.,
\etal 2004, ApJ, 617, 1272
\bibitem{} Fender, R. P., Belloni, T. \& Gallo, E. 2005, in From
X-ray Binaries to Quasars: Black Hole Accretion on All Mass
Scales, Eds. T. J. Maccarone, R. P. Fender and L. C. Ho, astro-ph
0506469
\bibitem{} Gallo, E., Fender, R. P., \& Pooley, G. G., 2003, MNRAS, 344, 60
\bibitem{} Halpern, J. P. 2005, ATEL 549
\bibitem{} Homan, J., Wijnands, R., van der Klis, M., \etal 2001, ApJS, 132, 377
\bibitem{} Homan, J. \& Belloni, T. 2005, Ap\&SS, 300, 107
\bibitem{} Kalemci, E., 2002, Bulletin of the American Astronomical Society, 201, 5705
\bibitem{} Maccarone, T. J. 2003, A\&A, 409, 697
\bibitem{} Magdziarz, P. \& Zdziarski, A. A. 1995, MNRAS, 273, 837
\bibitem{} Malzac, J., Petrucci, P.-O., Jourdain, E., et al. 2006,
A\&A, 448, 1125
\bibitem{} Markoff, S., Nowak, M. A. \& Wilms, J. 2005, ApJ, 635, 1203
\bibitem{} McClintock, J. E. \& Remillard, R. A. 2006, in "Compact Stellar
X-ray sources" (astro-ph 0306213)
\bibitem{} McConnell, M. L., Zdziarski, A. A., Bennett, K., \etal 2002, ApJ, 572, 984
\bibitem{} Miller, J. M., Homan, J. \& Miniutti, G. 2006, accepeted in ApJ (astro-ph 0605190)
\bibitem{} Mitsuda, K., Inoue, H., Koyama, K., \etal 1984, PASJ, 36, 741
\bibitem{} Morgan, E., Swank, J., Markwardt, C., \etal 2005, ATEL 550
\bibitem{} Morris, D. C., Burrows, D. N., Racusin, J., \etal 2005, ATEL 552
\bibitem{} Palmer, D. M., Barthelmey, S. D., Cummings, J. R., \etal 2005, ATEL 546
\bibitem{} Rodriguez, J., Corbel, S. \& Tomsick, J. A. 2003, ApJ, 595, 1032
\bibitem{} Rodriguez, J., Corbel, S., Kalemci, E., \etal 2004, ApJ, 612, 1018
\bibitem{} Still, M., Roming, P., Brocksopp, C., \etal 2005, ATEL 553
\bibitem{} Stirling, A., Spencer, R. E., de la Force, C. J., \etal 2001, MNRAS, 327, 1273
\bibitem{} Tanaka, Y., \& Shibazaki, N. 1996, ARA\&A, 34, 607
\bibitem{} Titarchuk, L.G. 1994, ApJ, 434, 570
\bibitem{} Torres, M. A. P., Steeghs, D., Garcia, M. R., \etal 2005a, ATEL 551
\bibitem{} Torres, M. A. P., Steeghs, D., Blake, C., \etal 2005b, ATEL 566

\end{thebibliography}
\end{document}